\documentclass[10pt,aps,nofootinbib,superscriptaddress,twocolumn,preprintnumbers,balancelastpage]{revtex4-1}
\usepackage{amsmath,amssymb,amsthm,amsopn}
\usepackage{epsfig}
\usepackage{psfrag}
\usepackage[colorlinks=true,allcolors=blue!65]{hyperref}
\usepackage{multirow}
\usepackage{empheq}
\usepackage{slashed}
\usepackage{graphicx}
\usepackage{url}
\usepackage{subfigure}
\usepackage{textcomp}
\usepackage{bm}
\usepackage{dcolumn}
\usepackage{color,xcolor}
\usepackage{ulem}
\usepackage{cancel}
\usepackage{braket}
\usepackage[title]{appendix}
\usepackage{lipsum}
\usepackage{mathptmx}

\DeclareMathAlphabet{\mathcal}{OMS}{cmsy}{m}{n}

\def\comment#1{}

\def\beq{\begin{equation}}
\def\eeq{\end{equation}}
\def\bea{\begin{eqnarray}}
\def\eea{\end{eqnarray}}

\begin{document}
\preprint{IPM/P-2021/007}

\title{Confronting the magnetically-induced holographic composite inflation with observation}

\author{M. Ahmadvand}
\email[]{ahmadvand@ipm.ir}
\affiliation{School of Physics, Institute for Research in Fundamental Sciences (IPM), P. O. Box 19395-5531, Tehran, Iran}

\author{A. Ashoorioon}
\email[]{amjad@ipm.ir}
\affiliation{School of Physics, Institute for Research in Fundamental Sciences (IPM), P. O. Box 19395-5531, Tehran, Iran}

\author{K. Bitaghsir Fadafan}
\email[]{bitaghsir@shahroodut.ac.ir}
\affiliation{Physics Faculty, Shahrood University of Technology, P.O.Box 3619995161 Shahrood, Iran}

\date{\today}

\begin{abstract}
We study the observational predictions of the phenomenological anti-de Sitter (AdS)/QCD inspired model, in which the inflaton field emerges in a four-dimensional strongly coupled gauge theory, in which the chiral symmetry breaking occurs through the formation of the quark condensate. Based on a top-down approach of AdS/QCD, using a D7-brane in the background of $N_c$ D3-branes, it has already been shown that chiral symmetry breaking  in a magnetic field through the generation of the Higgs vacuum expectation value could be a second-order phase transition, although it was doubted that this scenario could lead to enough inflation. Using an iterative method, we consistently solve for the  time-dependent parameters, including the embedding function of the D7-brane and  the Hubble parameter of the expanding background. We show that with $N_c\sim 10^7$ and $g_{\rm{}_{UV}}\sim {\rm few}\times 0.1$, the predictions of the inflationary model are consistent with the most stringent constraints placed on the inflationary models by Planck 2018. Although the model is capable of producing a large amount of gravitational waves, $r\simeq 0.01$,  the displacement of the canonical mass dimension-1 scalar field remains below the Planck mass, in violation of the Lyth bound.

\end{abstract}

\maketitle

\section{Introduction}
It is widely believed that in the very early stages after the big bang, the Universe  exponentially inflated. This accelerating period of expansion, which has been proposed as a solution for some issues such as the flatness and horizon problems \cite{Guth:1980zm,Linde:1981mu,Albrecht:1982wi}, can also serve as the origin of the present large-scale structures.

The physics of the inflation paradigm has been studied within various models \cite{Lyth:1998xn,Baumann:2009ds} and is able to satisfy the latest observational constraints \cite{Akrami:2018odb,Aghanim:2018eyx}. One of the well-motivated scenarios which may describe the inflation physics is strongly coupled gauge theories in which the inflaton field emerges as a bound state breaking a chiral symmetry at some energy scale \cite{Evans:2010tf,Channuie:2011rq} \footnote{In the context of strongly coupled gauge theory of chaotic inflation, see Refs.\ \cite{Harigaya:2012pg,Harigaya:2014wta}}. In fact, in this effective theory above this scale, there is no such bound state and instabilities of the inflaton mass, due to radiative corrections, can be circumvented \cite{Kachru:2003sx,McAllister:2005mq}.

In this paper, we investigate a strongly coupled gauge theory for inflation introduced in which chiral symmetry breaking occurs due to a background magnetic field \cite{Evans:2010tf}. In such theories, we cannot use conventional perturbative methods. One of the approaches through which many features of strongly coupled systems can be studied is anti-de Sitter (AdS)/CFT correspondence \cite{Maldacena:1997re,Witten:1998qj}. Employing this tool, the model is constructed from a D3/D7-brane system in the AdS/QCD framework \cite{Karch:2002sh,Erlich:2005qh}. In a phenomenological approach, the setup is exposed to a background magnetic field. It can be shown that the running of the gauge coupling is related to the magnetic field \cite{Filev:2007gb, Erdmenger:2007bn}, and indeed chiral symmetry breaking occurs through formation of the quark condensate in the IR energy scale. The time-dependent energy density of the condensate, which rolls slowly to the true vacuum, is calculated holographically.

In Ref.\ \cite{Evans:2010tf}, it was, however, doubted that the magnetic field induced by symmetry breaking, can yield a realistic model of inflation. The ansatz used for the background spacetime was an exponential scale factor with a constant Hubble parameter during the inflationary period. The potential for the condensate was derived holographically, but claimed to be too steep to sustain inflation. Therefore, Ref.\  \cite{Evans:2010tf} uses instead an ansatz for the running of the gauge coupling that resembles a step function. This way, the authors claim that they can flatten the potential around its maximum, elongating the inflationary period. However, here, we show that the magnetically induced symmetry breaking case can also cause a long enough inflationary period, which incidentally produces signatures compatible with the latest Planck results \cite{Akrami:2018odb}. In Ref.\ \cite{Evans:2010tf}, although the ansatz for the background spacetime is exponential, the energy density obtained from the embedding function is obviously time dependent, which is inconsistent with the initial assumption. Also, to demonstrate that the end of inflation is triggered with oscillations of the condensate around the vacuum, the Hubble parameter was separately taken to be zero; {\it i.e.}, the four-dimensional spacetime was taken to be flat.

To compare the background evolution with observation, we should simultaneously solve the background evolution with the Lagrange-Euler equation governing the embedding function. In this work, based on this phenomenological model, we consistently solve for the four-dimensional spacetime and the embedding function of the D7-brane, through an iterative method. This will allow us to see both the inflationary epoch and ensuing reheating phase in which the condensate oscillating phase, in a unified picture. In addition, we show that one can find microscopic parameters of the model such that the observables, including the amplitude of density perturbation, the scalar spectral index, and the tensor-to-scalar ratio, are all in agreement with the latest Planck measurement and bounds \cite{Akrami:2018odb}. We show that, although the model produces a large amount of tensor-to-scalar ratio, in violation of the Lyth bound \cite{Lyth:1996im}, the canonical mass dimension-1 scalar field displacement remains below the Planck mass.

The structure of the paper is as follows: In Sec.\ II, we review the string theory setup that realizes inflation and  try to iteratively solve for the background and the shape function until we approximately reach a fixed point. In Sec.\ III, we try to adjust the parameters such that the model is compatible with the latest Planck results. We conclude in Sec.\ IV.

\section{Holographic setup}\label{holo}

In the top-down approach to AdS/QCD, we construct D3/D7 setup in type IIB string theory, which corresponds to the large $N_c$ super-Yang-Mills $\mathrm{U}(N_c)$ gauge theory. In the higher-dimensional gravity side, D7-brane evolution as a probe in the background of $N_c$ D3-branes can be studied through Dirac-Born-Infeld (DBI) action. Here, the induced background is deformed in such a way that the desired inflationary mechanism in the four-dimensional dual field theory may be realized.

In the decoupling limit, we consider the $\mathrm{AdS}_5\times\mathrm{S}^5 $ solution by the metric \cite{Evans:2010tf}
\begin{align}	 ds^2&=g_{{}_{\mathrm{UV}}}^{-1}\Big[\frac{r^2}{R^2}\Big(g_{tt}dt^2+g_{ij}dx^idx^j\Big)\nonumber\\&+\frac{R^2}{r^2}\Big(d\rho^2+\rho^2d\Omega_3^2+dL^2+d\omega^2\Big)\Big]\,,
\end{align}
where $ g_{{}_{\mathrm{UV}}}$ is the asymptotic value of the gauge coupling, $ r^2=\rho^2+L^2$ (with $ \omega=0$, where  $\omega$ is the other coordinate perpendicular to the D7-brane)  is the radial coordinate corresponding to the energy scale of the gauge theory, $ R=(4\pi g_{{}_{\mathrm{UV}}}^2N_c\alpha'^2)^{1/4} $ is the AdS length scale, $ \alpha'\sim l_s^2$ is the string tension, $ l_s$ is the string length, $ g_{tt}=-1$, $ g_{ij}=a(t)^2\delta_{ij}$, and $ a(t)$ is the scale factor.

The effect of quarklike fields and their condensation which breaks the chiral symmetry of the theory is entered by considering the $ \mathrm{D}_7$-brane as well as $ \mathrm{D}_7$ embedding function, $ L(\rho,t)$.

According to the DBI action of the D7-brane \cite{Evans:2010tf}, the action for the embedding function is given by
\begin{equation}\label{dbi}
	S_{{}_{\mathrm{DBI}}}=-T_7R^4\int d^4xd\rho \frac{e^{\phi}~a(t)^3\rho^3}{\rho^2+L^2}\Gamma\,,
\end{equation}
where
\begin{equation}
	\Gamma\equiv\sqrt{-\dot{L}^2 +(\rho^2+L^2)^2(1+L'^2)},~~~~~~~T_7=\frac{1}{2(2\pi)^5g_{{}_{\mathrm{UV}}}^2\alpha'^4}\,,
\end{equation}
$ L'\equiv\partial_{\rho}L $, and $ \dot{L}\equiv\partial_{t}L$. Also, the running of the gauge coupling can be parametrized as
\begin{equation}
	e^{\phi}=g_{{}_{\mathrm{UV}}}^2\beta(r)=g_{{}_{\mathrm{YM}}}^2(r).
\end{equation}
The $\beta(r)$ function is related to the background magnetic field, $B$  \cite{Filev:2007gb,Jensen:2010ga,Evans:2010hi},
\begin{equation}
	\beta=\sqrt{1+\frac{B^2}{(\rho^2+L^2)^2}}\,,
\end{equation}
where $\beta(r)\rightarrow 1 $ when $r\rightarrow\infty $. From Eq.\ (\ref{dbi}), the equation of motion for the embedding function for the general four-dimensional background spacetime is obtained as follows
\begin{align}\label{el}
	\partial_{\rho}&\Big(\frac{\beta a^3\rho^3(\rho^2+L^2)L'}{\Gamma}\Big)+\partial_{t}\Big(\frac{\beta a^3\dot{L}}{(\rho^2+L^2)\Gamma}\Big)-\frac{2\Gamma a^3\rho^3L}{\rho^2+L^2}\frac{\partial\beta}{\partial r^2}\nonumber \\&-\frac{\beta a^3\rho^3}{\rho^2+L^2}\frac{\partial\Gamma}{\partial L}=0\,.
\end{align}
By finding $ L(\rho, t)$, one can obtain the four-dimensional time-dependent energy density from the energy-momentum tensor
\begin{equation}
	\langle T^{\mu\nu}\rangle=\frac{2}{\sqrt{-g}}\frac{\delta S_{{}_{\mathrm{DBI}}}}{\delta g_{\mu\nu}}\,,
\end{equation}
where $g_{\mu\nu} $ denotes the components of the boundary metric tensor. Therefore, the dimensionless energy density would be
\begin{equation}\label{T00}
	-T^0_0=\varepsilon(t)=\int d\rho\frac{\rho^3\beta}{\Gamma}(\rho^2+L^2)(1+L'^2)-\varepsilon_0\,,
\end{equation}
where $\varepsilon_0 $ is obtained from the case of static embedding function $ L_s\equiv L(\rho)$ by the following equation:
\begin{equation}\label{els}
	\partial_{\rho}\Big(\frac{\beta \rho^3L'}{\sqrt{1+L'^2}}\Big)-2\rho^3L\sqrt{1+L'^2}\frac{\partial\beta}{\partial r^2}=0\,.
\end{equation}
For the dimensionful energy density $\varepsilon$ in Eq. \eqref{T00}  should be multiplied with $T_7 R^4 $.

\begin{figure}
	\includegraphics[scale=0.6]{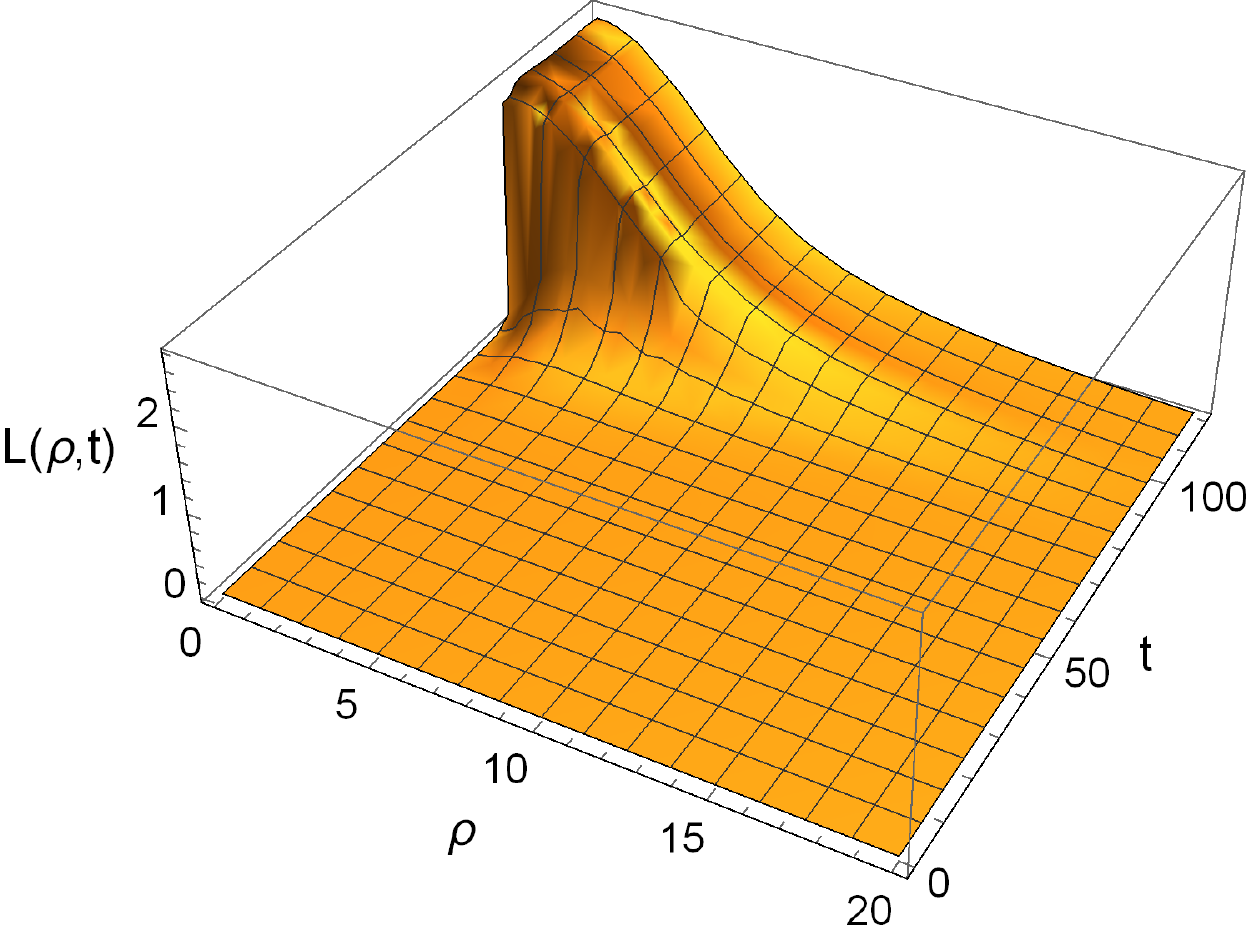}
	\caption{Using the iterative approach and solving Eq.\ (\ref{el}), the embedding function $ L(\rho,t)$ representing the transition between vacua of the model is obtained.}
	\label{f-1}
\end{figure}

\begin{figure}
	\includegraphics[scale=0.37]{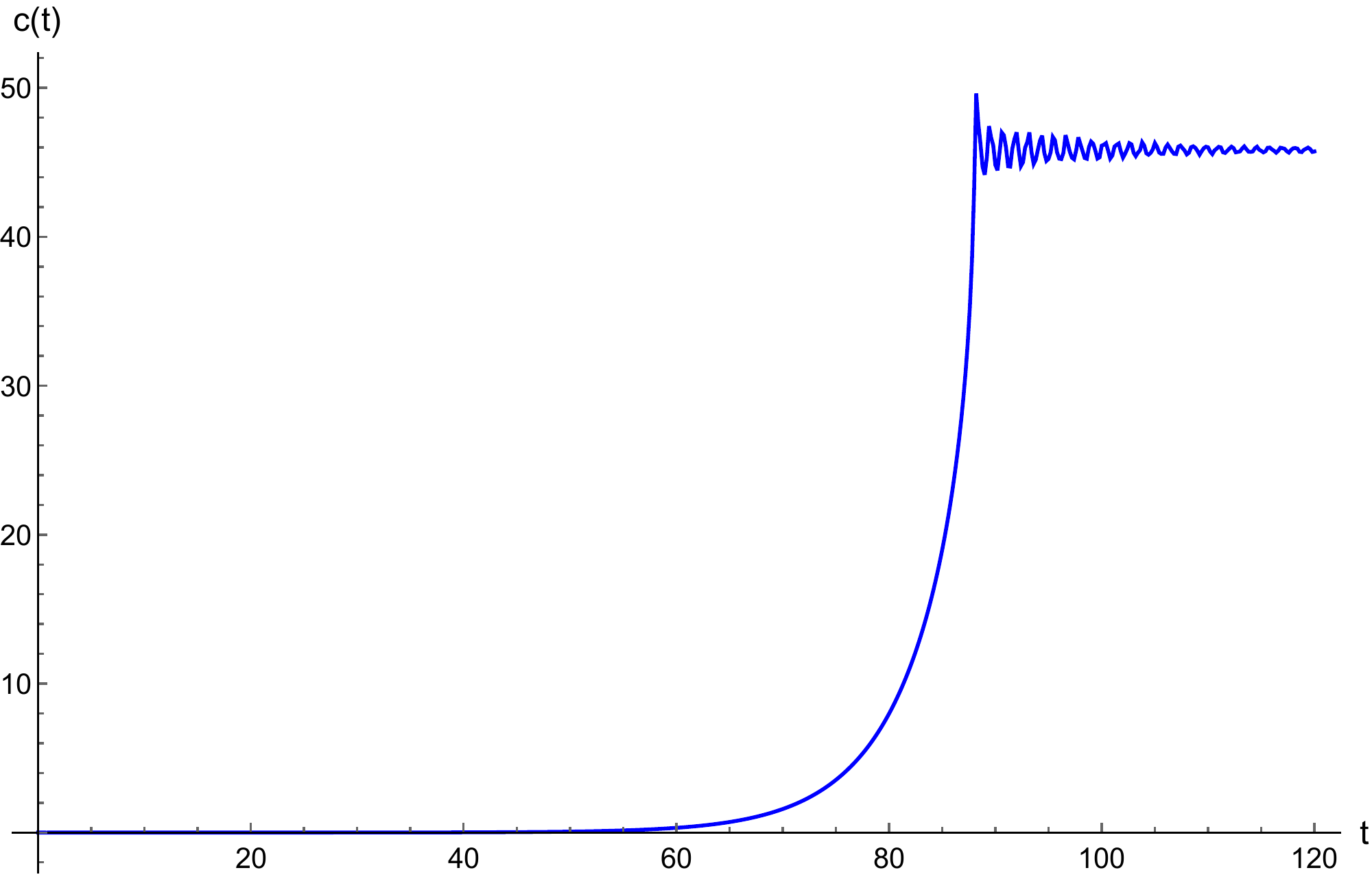}
	\caption{The time evolution of the condensate is shown. As expected, it oscillates around the true vacuum at late time.}
	\label{f0}
\end{figure}

\begin{figure}
	\includegraphics[scale=0.4]{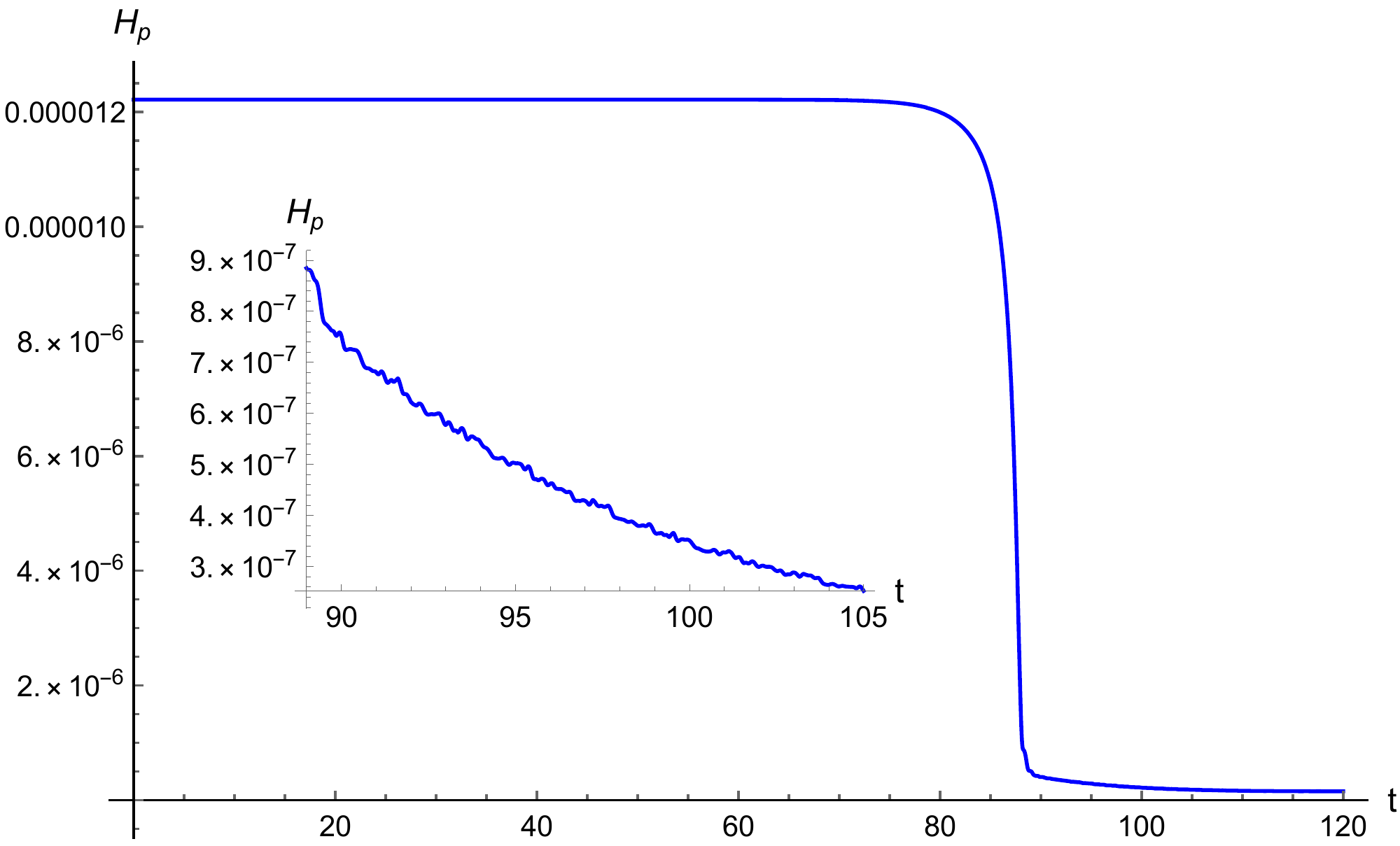}
	\caption{The Hubble parameter is displayed for $ N_c=10^7$, $ \lambda=1.4\times 10^6$, and $ n_p=8\times 10^5$. As expected, the Hubble parameter has a decreasing oscillatory behavior at late times, after inflation ends.}
	\label{f1}
\end{figure}

To obtain the time-dependent energy density and then the Hubble parameter consistently, $ H(t)=\sqrt{\varepsilon(t)/3}$, we use a numerical iterative approach. For the initial ansatz for the four-dimensional spacetime and as we look for a quasiexponential expanding background, we take the Hubble parameter at first order to be constant, $ H^{(0)}$, which in turn implies $ a^{(0)}(t)\sim \exp (H^{(0)}t)$. Imposing appropriate initial conditions, we can numerically find the embedding function at this order and, in the next step, determine the the energy density from \eqref{T00}. Having the energy density, we again determine the resulting Hubble parameter from the first Friedmann equation. These steps can be iterated as many times as needed, until one converges on a fixed point which simultaneously solves both equations \eqref{T00} and the Friedmann equation. Following Ref.\ \cite{Evans:2010hi}, we take the representative values for  $ H^{(0)}\simeq 4.83$, $ B=35.6$. Also, to ensure the  slow roll for the evolution of the condensate at the false vacuum, we consider the initial condition $ L(\rho, 0)=0$, $ \dot{L}(\rho, 0)=10^{-5} \exp (-\rho^2)$, and $ L^{\prime}(0, t)=L(\infty, t)=0$ as the boundary condition. Solving for $L^{(0)}(\rho, t)$, one can obtain the energy density from \eqref{T00} and then $H^{(1)}(t) $ from the first Friedmann equation. Plugging $ a^{(1)}(t)\simeq \exp (\int dt H^{(1)} )$ again into Eq.\ \eqref{el}, we can obtain the embedding function and then $H^{(2)}\equiv H(t)$. We have iterated this procedure one more time, up to $L^{(2)}$, and made sure of the smallness of the correction at this order, by verifying that
\begin{equation}\label{convergence}
    \frac{L^{(2)}-L^{(1)}}{L^{(1)}}<1\,.
\end{equation}
 Using the iterative approach up to the second order, which suffices to show expected properties, the embedding function is numerically found as shown in Fig.\ \ref{f-1}.
UV asymptotic embedding solution corresponds to the function \cite{Babington:2003vm}
\begin{equation}\label{dual}
	L(\rho,t)=m+\frac{c(t)}{\rho^2},
\end{equation}
where $ m$ is proportional to the quarklike mass, $ m_q=m/(2\pi\alpha')$ and $ c(t)$ is associated to the condensate $ \langle\bar{q}q\rangle (t)=c(t)/(2\pi\alpha')^3 $ \cite{Erdmenger:2007bn}. Note that we worked with dimensionless variables of ($ L,\rho, m, c$) and in order to be dimensionful $ L,\rho$, and  $m$ should be rescaled by $ R$ and $ c$ by $ R^3$. Based on the obtained $ L(\rho,t)$ and its numerical fit to (\ref{dual}) in a small range in large $ \rho$, we can extract $ m$ and $ c(t)$ and figure out the composite inflaton dynamics through $ c(t)$. Interestingly, we obtain not only the rolling phase of the condensate but also its oscillatory behavior around its true vacuum at late times, as it is  evident from Fig.\ \ref{f0}. The oscillating phase corresponds to the time when inflation has ended and the Universe starts decelerating again, leading to matter- or radiation-dominated universe.
To contrast the parameters with observable values, we take into account dimensionful parameters in Planck units. Thereby, the energy density in Planck units is defined as
\begin{equation}\label{pl}
	\varepsilon_p\equiv T_7 R^4\varepsilon=M_p^4\tilde{\varepsilon}\,,
\end{equation}
where the Planck mass can be related to the AdS length scale as $ M_p=n_p/R$ and $ n_p$ as a number will be fixed in the following. Therefore, from Eq.\ (\ref{pl}),
\begin{equation}
	\tilde{\varepsilon}=\frac{2\lambda N_c}{(2\pi)^3n_p^4}\varepsilon\,,~~~~~~~~\lambda=g_{{}_{\mathrm{UV}}}^2N_c\,,
\end{equation}
and the Hubble parameter will be
\begin{equation}\label{hp}
	H_p=\sqrt{\frac{8\pi\varepsilon_p}{3M_p^2}}=\sqrt{\frac{16\pi\lambda N_c\varepsilon}{3(2\pi)^3n_p^2R^2}}\,.
\end{equation}
Eventually, by finding $ \varepsilon(t)$ based on the aforementioned approach and fixing the variables in Eq.\ (\ref{hp}) based on observable slow-roll parameters, discussed below, the Hubble parameter can be obtained. We show the result per Planck mass for $ M_p=1$, {\it i.e.}, $ R=n_p$ in Fig.\ \ref{f1}. In addition, we display the oscillatory behavior of the Hubble parameter at late times after inflation ends.
%Since we are dealing with derivative and second derivative of the Hubble parameter in the next section, to avoid numerical errors, we find the best fit function which is given in Appendix \ref{fit}.
\section{Holographic Inflation as an Effective single-field inflation}
Since we consider a holographic setup with extra dimensions, in principle quantum fluctuations of D7 embedding function in these degrees of freedom may play a role in the inflationary dynamics. However, in this section, we show that with some assumptions, the treatment of the setup effectively as single-field inflation is correct. The $L$ fluctuations can be expressed as
\begin{equation}
\delta L(\rho, t)=\frac{\partial L/\partial t}{\partial c/\partial t}\delta c+\frac{\partial L}{\partial \rho}\delta\rho\equiv a_c \delta c+a_{\rho}\delta\rho\,.
\end{equation}
The power spectrum of $\langle \delta L\delta L\rangle $ is related to  the power spectra of these fluctuations through the relation,
\begin{equation}\label{fluc}
\langle \delta L\delta L\rangle=a_c^2\langle \delta c \delta c \rangle+a_{\rho}^2\langle \delta\rho \delta\rho \rangle
\end{equation}
\begin{figure}
	\includegraphics[scale=0.53]{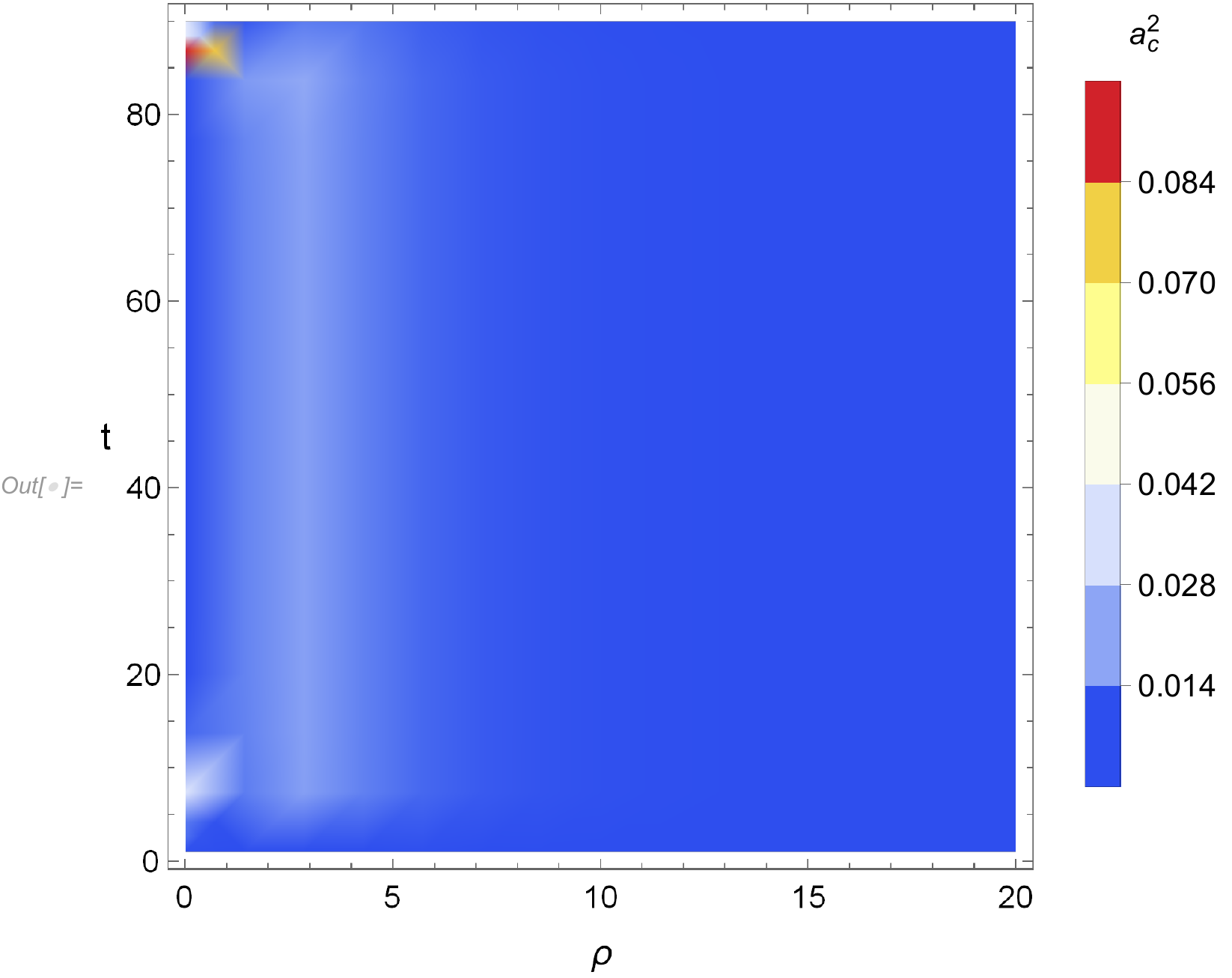}
	\caption{The $ a_c^2$ factor is plotted. The fluctuations are suppressed by this factor within the parameter values.}
	\label{fd1}
\end{figure}
\begin{figure}
	\includegraphics[scale=0.53]{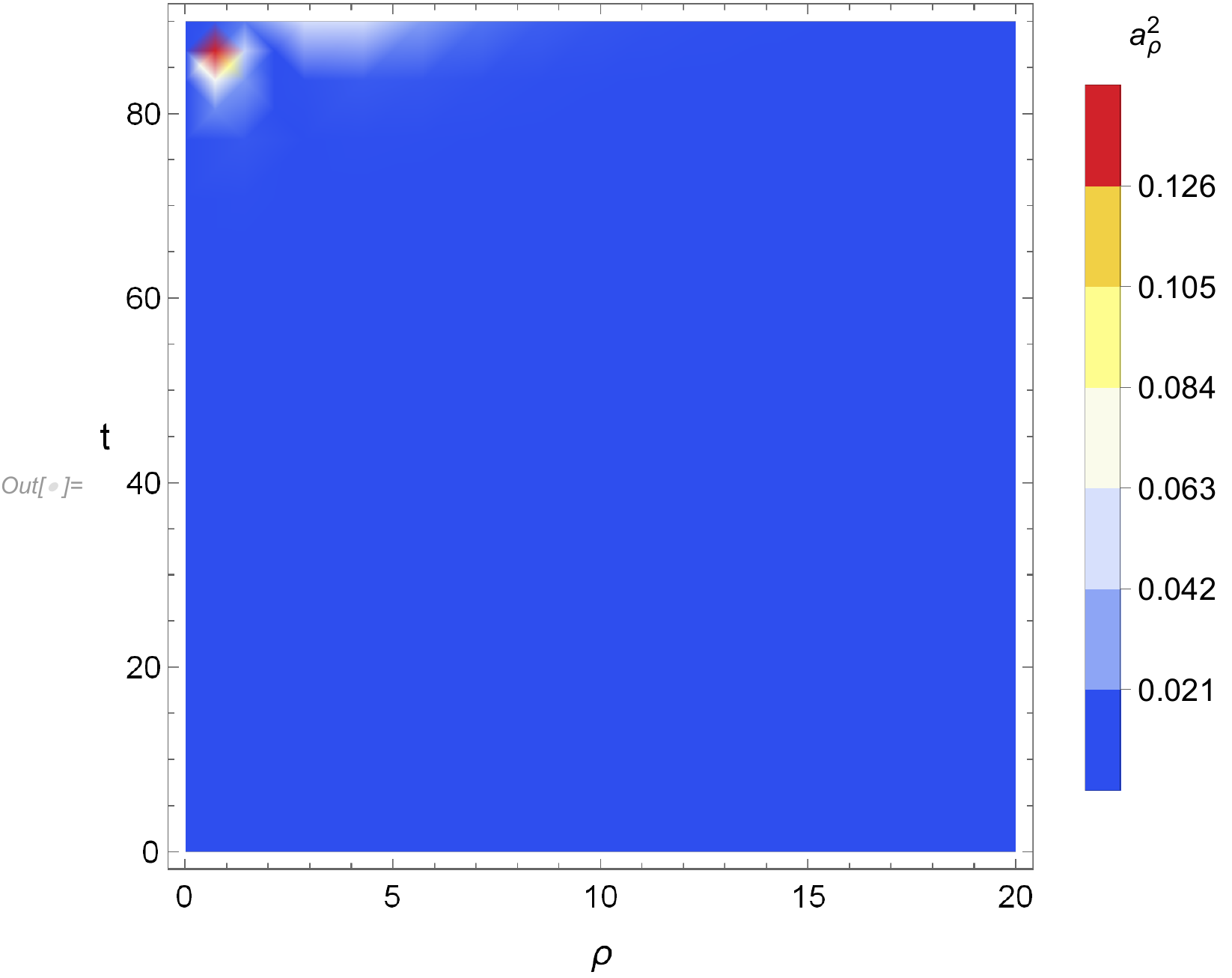}
	\caption{The $ a_{\rho}^2$ factor is plotted within the parameter values.}
	\label{fd2}
\end{figure}
To prove that our single-field treatment is justified, we have to show that the spectra of other fields are subdominant with respect to the quark condensate, $c$, fluctuations. As can be seen from Fig.\ \ref{fd1}, we find that the power spectrum of $L$ fluctuations is suppressed relative to the $c$ perturbations by the factor $a_c^2$ , where $ a_c^2$ is at most a factor of a percent at the cosmic microwave background (CMB) scales. Despite the smallness of $a_{\rho}^2$, as shown in Fig.\ \ref{fd2}, we also have to assume that in the full-fledged theory the mass parameter associated with the $\delta \rho$ fluctuations, which represent the moduli of the D7-brane in four other extra dimensions, is stabilized and thus sufficiently heavy to neglect the second term of Eq.\ (\ref{fluc}) in comparison with the first one. \footnote{Assuming that $\rho$ modulus is stabilized, the trajectory in the $\rho -c$ plane is a straight line. Hence $\delta c$ and $\delta \rho$ fluctuations are not cross-correlated.}.

In the next section, we study whether the predictions of this effective model are compatible with observation.

\section{Slow-roll inflation}
To describe the slow-roll inflation, we should investigate the conditions which can be achieved by the time-dependent Hubble parameter in the model. The first slow-roll parameter, quantifying whether inflation takes place at a given time, is written as
\begin{equation}
	\epsilon\equiv\frac{-\dot{H}}{H^2}\,.
\end{equation}
Note that in Planck units the dimensionless time should be replaced, $ t\rightarrow n_p t$, for $ n_p=R$ and hence $\epsilon=-\dot{H_p}/(n_pH_p^2) $. During the slow-roll inflation, $ \epsilon<1$. Inflation also ends when $ \epsilon(t_{\mathrm{end}})=1$.

To maintain inflation for sufficient amount of time, the second condition, which should be less than 1 during this period, is given by
\begin{equation}
	\eta\equiv\epsilon-\frac{\dot{\epsilon}}{2\epsilon H}=\epsilon-\frac{\dot{\epsilon}}{2n_p\epsilon H_p}\,.
\end{equation}
The amount of inflation can be quantified in terms of the number of $e$-folds as
\begin{eqnarray}
	N_e\equiv\int_{t_{{}_{\rm CMB}}}^{t_{{}_{\mathrm{end}}}}dt~H=\int_{t_{{}_{\rm CMB}}}^{t_{{}_{\mathrm{end}}}}dt~n_p H_p\,.
\end{eqnarray}
To solve the big bang problems, for the grand unified theory (GUT-) scale inflation, one needs inflation to last at least $ N_e\simeq 60$, before it ends at $t_{{}_\mathrm{end}}$. In the model, satisfying the slow-roll conditions and other observables that will be discussed, this can be fulfilled for, e.g., $ t_{{}_{\rm CMB}}=80.6646$ and $ t_{{}_{\mathrm{end}}}=89.6367$.

\begin{figure}
	\includegraphics[scale=0.65]{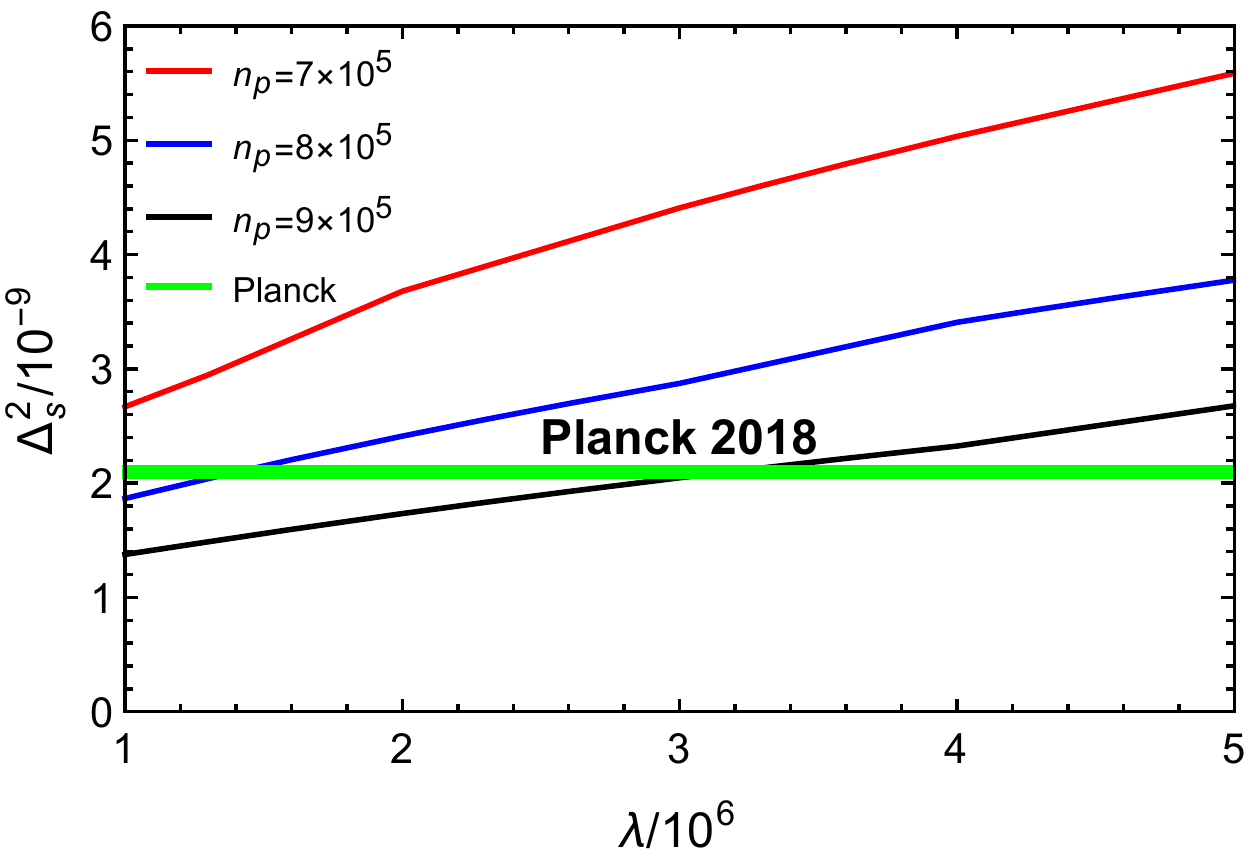}
	\caption{For three different values of $ n_p$, the magnitude of scalar power spectrum is displayed vs $ \lambda$ for $ N_c=10^7$ and $ N_e=60$.}
	\label{f2}
\end{figure}

\begin{figure}
	\includegraphics[scale=0.65]{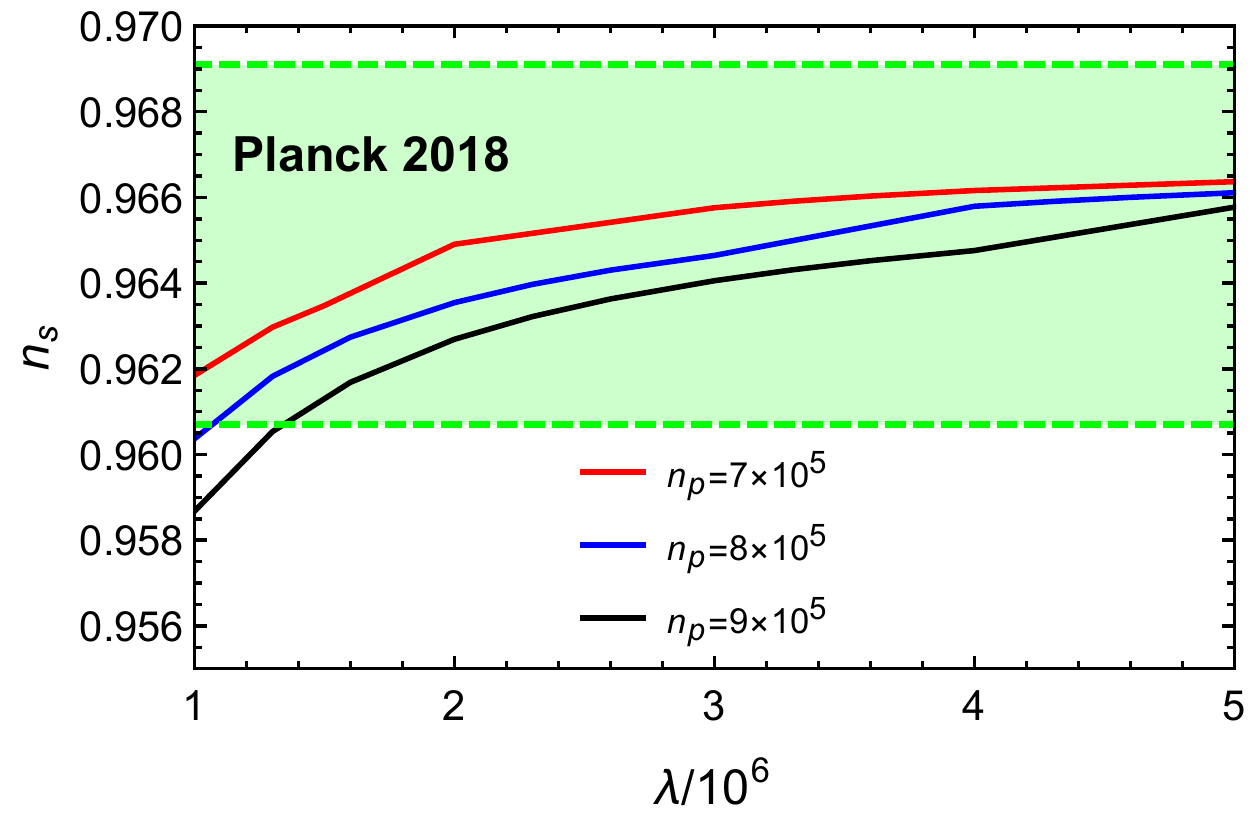}
	\caption{For three different values of $ n_p$, the scalar spectral index is displayed vs $ \lambda$ for $ N_c=10^7$ and $ N_e=60$. The green region determines the valid range for $ n_p$.}
	\label{f3}
\end{figure}

\begin{figure}
	\centering
	\includegraphics[scale=0.65]{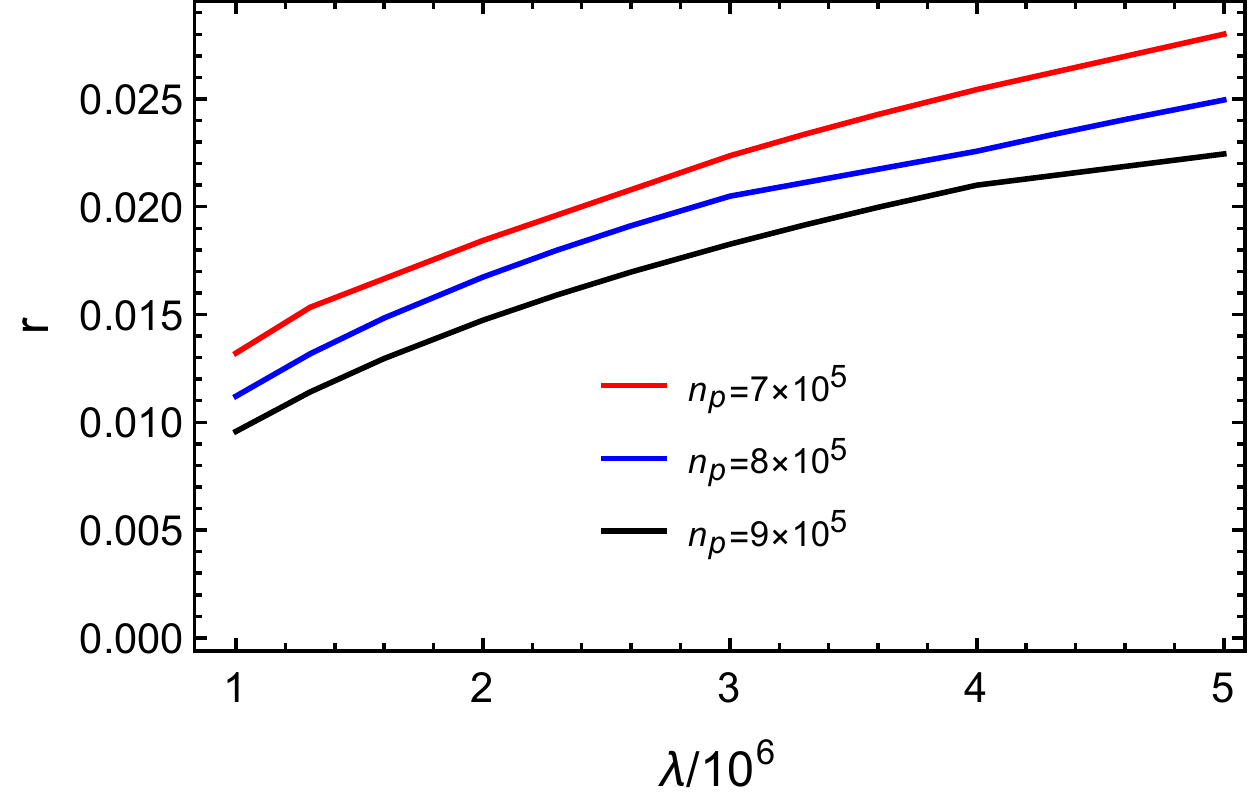}
	\caption{Compatible with the upper limit $r<0.056 $, the tensor-to-scalar ratio is plotted vs $ \lambda$, for three different values of $ n_p$, $ N_c=10^7$ and $ N_e=60$.}
	\label{f4}
\end{figure}

Now, we can proceed to calculate relevant observable parameters to study the inflationary model. During inflation, scalar and tensor perturbations can induce fluctuations in the CMB. The dimensionless power spectrum of scalar fluctuations is defined as
\begin{equation}
	\Delta_s^2\equiv\frac{1}{8\pi^2}\frac{H_p^2}{\epsilon}\,.
\end{equation}
To explain the large-scale structures due to these fluctuations, the magnitude is required to be in the range of $\ln (10^{10}\Delta_s^2)=3.044\pm 0.014 $ at $ t_{{}_{\rm CMB}}$ obtained from Planck temperature, polarization, and lensing data \cite{Akrami:2018odb,Aghanim:2018eyx}. The deviation of the scalar power spectrum from scale invariance can be parametrized in terms of the scalar spectral index, which is defined in terms of slow-roll parameters as
\begin{equation}
	n_s=1+2\eta-4\epsilon\,.
\end{equation}
$ n_s$ is calculated at $t_{{}_{\rm CMB}}$ and the observational constraint on this parameter is $ n_s=0.9649\pm 0.0042 $ \cite{Akrami:2018odb,Aghanim:2018eyx}. Another important parameter is the tensor-to-scalar ratio, which is indeed a measure of magnitude of the tensor power spectrum, $ r\equiv\Delta_t^2/\Delta_s^2 $. The tensor power spectrum is proportional to the Hubble parameter squared \cite{Lidsey:1995np}. This is true even in the presence of extra dimensions \cite{Maartens:1999hf}. Hence, we can obtain the tensor-to-scalar ratio at $t_{{}_{\rm CMB}}$ from the first slow-roll parameter
\begin{equation}
	r\simeq 16\epsilon\,.
\end{equation}
The current upper limit on $ r$ is $ r<0.056$ \cite{Akrami:2018odb,Aghanim:2018eyx}. Within the parameter space of the model, we can find the observable parameters in the valid range, with adjusting the parameters in the microscopic theory. For example, for $ N_c=10^7$, $ \lambda=1.4\times 10^6$, i.e. $ g_{\mathrm{UV}}\simeq 0.374$, $ n_p=8\times 10^5$, and $ N_e=60$ at $ t_{{}_{\rm CMB}}$, we obtain the magnitude of scalar power spectrum $ \Delta_s^2=2.09539\times 10^{-9}$, the scalar spectral index $ n_s=0.962171$ and then $ r=0.0137587$. Furthermore, based on the numerical analysis and fixed values of variables shown in Figs.\ \ref{f2}, \ref{f3}, and \ref{f4}, we display a set of values of $ \Delta_s^2$, $ n_s$, and $ r$ for 60  $e$-folds and different values of $ \lambda$ \footnote{Matrix inflation is another string theory motivated model with a very large gauge group \cite{Ashoorioon:2009wa,Ashoorioon:2011ki}. The number of D3-branes could be reduced substantially, if the inflaton is nonminimally coupled to gravity \cite{Ashoorioon:2019kcy}.}.

We have checked that in the next orders of iteration observables may be fixed as $ \Delta_s^2\sim2.09\times 10^{-9}$, $ n_s\sim0.96$, and $ r\sim0.01$ for $ N_e=60$ provided that at the given order, the parameters, e.g., the coupling $\lambda$, change subsequently. As a result, the proposed inflationary model is able to provide a viable range of observable parameters in agreement with experimental measurements and bounds.


\subsection{Lyth bound}

In the previous section, we found the tensor-to-scalar ratio $ r\sim0.01 $. This implies that inflation takes place at GUT energy scales $ (r/0.01)^{1/4} 10^{16}\,\mathrm{GeV}$. The Lyth bound relates $r$ to the inflaton field excursion during inflation \cite{Lyth:1996im},
\begin{equation}
	\frac{\Delta \phi}{M_p} \simeq \mathcal{O}(1)\left(\frac{r}{0.01}\right)^{1 / 2}\,,
\end{equation}
 and hence for the obtained $ r\gtrsim 0.01$ of the model, one would expect a large field model of inflation. We now try to verify whether this bound is satisfied in our model. One should note that the condensate is a mass dimension cubed composite scalar  field,
 \begin{equation}
\langle\bar{q}q\rangle = R^3 c/(2\pi\alpha')^3\,,
\end{equation}
and in order to transform it to a canonically normalized mass scalar field, with mass dimension equal to one, it should be divided by a mass dimension squared parameter,
\begin{equation}
	\Delta \phi\sim \frac{R^3 \Delta c}{(2\pi\alpha')^3 \Lambda^2}\,,
\end{equation}
where $ \Lambda$ is the cutoff of the theory.  On the other hand, the UV cutoff of the theory should be greater than the symmetry breaking scale $ \Lambda>R\,c^{1/3}/(2\pi\alpha')$, In Planck units where $ R=n_p$ and hence $ l_s\sim 10^4$, $\Delta \phi\lesssim 10^{-3} $. Therefore, despite predicting $r\gtrsim 0.01$, the excursions of the canonical field remain below $M_p$.

\section{Conclusion}

Trying to find a mechanism within the fundamental theories for inflation is still an interesting arena. One can gain substantial constraints and information about these theories. In particular since the upper bound on the tensor-to-scalar ratio $r<0.056$ is higher than the borderline value $r_c\approx 0.01$, which divides the small and large single-field models in the single-field inflation, obtaining $r\gtrsim 0.01$ with excursions below $M_p$ for the canonical field, which controls the effect of higher-dimensional operators is of interest.

In this paper, we have explored a phenomenological holographic model constructed from the D3/D7-brane system in a background magnetic field. Based on the DBI action, where deformation to the gauge coupling is produced due to the magnetic field, a phenomenological slow-roll inflationary model had been constructed \cite{Evans:2010tf}. However it was doubted that the model is even able to produce long enough inflationary period that solves the problems of big bang cosmology. Using an iterative approach, we have numerically obtained the embedding function related to the condensate and shown its oscillatory behavior at late time around the true vacuum, in addition to its rolling toward it in the beginning. We demonstrate that the model is not only able to produce long enough inflation, but can be adjusted to produce the scalar power spectrum and  scalar spectral index compatible with the latest Planck 2018 constraints. Eventually, we showed that, despite the model being able to produce $r\gtrsim0.01$,  the excursions of the canonical field remain below $M_P$.

It will be interesting to study the model further, particularly in connection to the mechanism(s) for reheating after inflation.

\section{acknowledgments}
The work is supported by Iran National Science Foundation (INSF).

%\begin{appendices}
%	\section{Fit function}\label{fit}	
%	In this appendix, we introduce the fit function of the Hubble parameter obtained numerically in Section \ref{holo}. Using the Hubble parameter data, Fig.\ (\ref{f1}), we find the function which is greatly matched to it. The function is as follows
%	\begin{align}
%		H(t)&=A+\frac{B \left(\left(\frac{-\frac{F Q}{2 U}-G+t}{Q}\right)^2+1\right)^{-U} }{\left(\frac{F^2}{4 U^2}+1\right)^{-U}}\times\nonumber\\
%		&\times \exp \left(-F \left(\tan ^{-1}\left(\frac{-\frac{F Q}{2 U}-G+t}{Q}\right)+\tan ^{-1}\left(\frac{F}{2 U}\right)\right)\right).
%	\end{align}
%	The following figure of the fit function can be displayed if the coefficients are chosen as: $ A=-0.01672630$, $ B=4.67710351$, $ G=65.12989115$, $ Q=0.511094276$, $ U=0.014162465$, and $ F=1.266533393$.
%	\begin{figure}
%		\includegraphics[scale=0.7]{p5.pdf}
%		\caption{For $ N_c=10^7$, $ \lambda=1.4\times 10^6$ and $ n_p=8\times 10^5$, the Hubble parameter data, the blue curve, and its fit function, the red curve, are plotted.}
%		\label{f5}
%	\end{figure}
%\end{appendices}

\end{document}